\documentclass[aps,twocolumn, showpacs,superscriptaddress,floatfix]{revtex4}
\usepackage{graphicx}
\usepackage{amsmath,bbm,amssymb}

\def\Tr{{\rm Tr}}
\newcommand{\n}{\mbox{\boldmath $\nabla$}}

\newcommand{\Zb}{{\bf Z}}
\newcommand{\Kb}{{\bf K}}

\newcommand{\ep}{\varepsilon}

\newcommand{\res}{C_{\rm res}}

\begin{document}
\bibliographystyle{prsty}
\title{Relaxation of a qubit measured by a driven Duffing oscillator}
\author{I.~Serban}
\affiliation{IQC and Dept.~of Physics and Astronomy, University of Waterloo, 200 University Ave W, Waterloo, ON, N2L 3G1, Canada}
\affiliation{Department Physik, Arnold-Sommerfeld-Center for Theoretical Physics, and Center for Nanoscience, Ludwig-Maximilians-Universit\"at, Theresienstr.~37, 80333 M\"unchen, Germany}
\affiliation{Instituut Lorentz, Leiden University, Niels Bohrweg 2, NL-2333 CA Leiden,
The Netherlands}
\author{M. I.~Dykman}
\email{dykman@pa.msu.edu}
\affiliation{Department of Physics and Astronomy,
Michigan State University, East Lansing, MI 48824, USA}
\author{F. K.~Wilhelm}
\email{fwilhelm@iqc.ca}
\affiliation{IQC and Dept.~of Physics and Astronomy, University of Waterloo, 200 University Ave W, Waterloo, ON, N2L 3G1, Canada}

\date{\today}
\begin{abstract}
We investigate the relaxation of a superconducting qubit for the case when its detector, the Josephson bifurcation amplifier, remains latched in one of its two (meta)stable states of forced vibrations.
The qubit relaxation rates are different in different states. They can display strong dependence on the qubit frequency and resonant enhancement, which is due to
{\em quasienergy resonances}. Coupling to the driven oscillator changes the effective temperature of the qubit.
\end{abstract}

\maketitle

\section{Introduction}
The Duffing oscillator is a paradigmatic model for nonlinear dynamics. Its quantum properties \cite{Larsen1976,Sazonov1976,Drummond1980c,Dmitriev1986a,Dykman1988a,Peano2006,Dykman2007,Serban2007}  have gained renewed recent
interest in the light of nanomechanics \cite{Kozinsky07} and as a qubit readout device \cite{Siddiqi04,Siddiqi05,Lupascu07,Metcalfe07} for superconducting qubits \cite{Clarke2008}, the Josephson bifurcation amplifier (JBA). In the context of readout, examples of the experimental setup are a driven nonlinear superconducting resonator \cite{Metcalfe07} coupled to the quantronium qubit \cite{Vion02} or a superconducting quantum interference device (SQUID) used as a JBA \cite{Lupascu07}, inductively coupled to the flux qubit and in contact with dissipative measurement circuitry.

The JBA detection profits from the very different amplitudes and phases of the coexisting states of oscillator's forced vibrations and the formally infinite gain at the bifurcation \cite{Siddiqi04,Siddiqi05,Lupascu07,Metcalfe07}. In the first step of detection, the JBA is driven through the bifurcation point where, depending on the state of the qubit, one of the attractors disappears. The JBA then decides between the attractors thus pre-measuring the qubit. This decision is probabilistic, because of quantum and classical fluctuations that can cause an inter-attractor transition even before the bifurcation point is reached. At the next stage, the parameters are changed so as to suppress the probability of fluctuation-induced switching and the latched vibration state is observed. Qubit energy relaxation in the latter state will drive the measurement away from the ideal quantum nondemolition regime in that the post-measurement state is not correlated with the measurement output \cite{Picot08}.

In this paper, we study qubit relaxation mediated by the
oscillator. Because of the oscillator-qubit interaction, coupling of the oscillator to a heat bath leads to an indirect coupling of the qubit to that same bath and, consequently, to relaxation. The energy emitted by the qubit creates excitations in the bath. The qubit transition can be {\it stimulated} by forced vibrations of the oscillator, which have a comparatively large amplitude already for a weak resonant driving. This means that the driving accelerates qubit relaxation. For the lowest-order stimulated processes that we consider, the energy transfer to the bath becomes $|\omega_q\pm \omega_F|$, where $\omega_q$ is the qubit frequency and $\omega_F$ is the driving frequency. As a consequence, the qubit relaxation rate becomes dependent on the vibration amplitude of the oscillator, which bears on the experimental observations \cite{Picot08}.

The dependence on the vibration amplitude is even more pronounced where $\omega_q/2$ is close to the oscillator eigenfrequency $\omega_0$, $|\omega_q-2\omega_0|\ll \omega_0$. In this case the decay rate displays resonances where $\hbar(\omega_q-2\omega_0)$ goes through the quasienergy level spacing of the driven oscillator. We have analyzed both resonant and nonresonant cases. The results coincide where the ranges of their applicability overlap, which indicates that they describe all qubit frequencies. We have also extended the analysis to the case where $\omega_q$ is close to $\omega_0$.

In Sec.~II we give a general expression for the relaxation rate of a qubit coupled to a driven quantum oscillator. In Sec.~III we describe the theory of the oscillator and its fluctuations close to the stable vibrational states. In Sec.~IV we provide an explicit expression for the oscillator power spectrum. The proposed method and the result go beyond the immediate problem of the qubit relaxation. The obtained expression allows us to study the qubit relaxation rate as a function of frequency detuning $\omega_q-2\omega_F$, including the onset of quasienergy resonances, and to discuss resonant heating and cooling of the qubit. In Sec.~V qubit relaxation and the change of the effective qubit temperature are studied far from resonance. Section~VI provides a summary of the results.

\section{Golden rule qubit relaxation}

We will express the qubit relaxation rate in terms of the power spectra of small-amplitude fluctuations of the driven oscillator about its steady states of forced vibrations. The fluctuations will be assumed small, which justifies describing them in a
linearized approximation. This approach allows us to find the relaxation rate both where the qubit frequency $\omega_q$ is close or far detuned from the oscillator eigenfrequency $\omega_0$ and $2\omega_0$.

The Hamiltonian of a superconducting qubit reads
\begin{equation}
\label{eq:qubit_Hamiltonian}
\hat H_{\rm q}=\hbar \frac{w}{2}\hat\sigma_z +\hbar\frac{\delta}{2}\hat\sigma_x,
\end{equation}
where $\sigma_{x,y,z}$ are Pauli matrices and $|\delta|\ll w$. The inductive coupling of the qubit to the oscillator will be treated as a weak perturbation and, for the typical setup of superconducting qubit measurements, is quadratic in the oscillator coordinate $x$ \cite{Siddiqi05,Lupascu07,Siddiqi2006,SerbanPRB07,Lupascu04,Ioana07a,SerbanPRB07}
\begin{equation}
\hat H_{I{\rm d}}=\frac{m\omega_0\Delta_q}{2}\hat\sigma_z\hat x^2,
\label{eq:couplinghamiltonian}
\end{equation}
where $\Delta_q$ is the oscillator frequency shift  due to the coupling. We will also discuss the linear case, in which the operator $x^2$ in Eq.~(\ref{eq:couplinghamiltonian}) is replaced by $x$. As the interaction (\ref{eq:couplinghamiltonian}) does not commute with the qubit Hamiltonian $\hat H_{\rm q}$ it introduces energy relaxation.

We assume that the oscillator is coupled to the bath strongly enough, so that its damping $\kappa$ exceeds the decay rate of the qubit. Then the oscillator serves as a bath for the qubit. To second order in $\Delta_q$, the decay rate of the qubit excited state $\Gamma_e$ has a standard golden rule form \cite{Blum96,Chemphys04,Nato06II,DK_review84,Shnirman2005},
\begin{eqnarray}
\label{eq:decay_excited_general}
\Gamma_e= C_{\Gamma}{\rm Re}\,G(\omega_q),\qquad C_{\Gamma}=\frac{1}{2}(m\omega_0\Delta_q\delta/\hbar\omega_q)^2,\\
G(\omega)=\int\nolimits_0^{\infty}dt e^{i\omega t}\langle\delta \hat x^2(t+\tau)\delta \hat x^2(\tau)\rangle_F,\nonumber
\end{eqnarray}
where $\omega_q=\sqrt{w^2+\delta^2}$ is the qubit transition frequency and $\delta \hat x^2(t)\equiv \hat x^2(t)-\langle \hat x^2(t)\rangle$.
The averaging $\langle\cdot\rangle$ is done disregarding the qubit-oscillator interaction and assuming that the oscillator {\it fluctuates about a given steady state of forced vibrations}. In calculating $\langle \hat A(t+\tau)\hat B(\tau)\rangle_F$ we imply additional averaging with respect to $\tau$ over the driving period; this period is small compared to
 $\Gamma_e^{-1}$ due to the underdamped nature of the system.

The expression for the excitation rate out of the qubit ground state $\Gamma_g$ has the same form as $\Gamma_e$ except $G(\omega_q)$ should be replaced by $G(-\omega_q)$.
We will see that the ratio $\Gamma_g/\Gamma_e$ is not given by the Boltzmann factor $\exp(-\hbar\omega_q/k_BT)$ due to the nonequilibrium character of this driven system. The Bloch-Redfield rates for the qubit  are
\begin{eqnarray}
\label{eq:Bloch_rates}
&&T_1^{-1}=\Gamma_e+\Gamma_g, \\
&&T_2^{-1}=\frac{1}{2}T_1^{-1}+ 2C_{\Gamma}(w/\delta)^2{\rm Re}\,G(0).\nonumber
\end{eqnarray}
Of primary interest for us will be the decay rate $T_1^{-1}$. As we show, it has a component where one quantum of the oscillator or the bath is created or annihilated; this component is proportional to the squared dimensionless amplitude
$r_a^2$ [see Eq.~\ref{eq:equilibrium_position})] of the attractor the JBA has latched to; $T_1^{-1}$ also has a two-quantum component. Both of them can lead to enhanced relaxation at specific working points, identified as {\em quasienergy resonances}.

\noindent
\section{Dynamics of a resonantly driven nonlinear oscillator}
To find the oscillator correlation function $G(\omega)$ we assume the limit where the resonant driving current $I_B$ of the corresponding SQUID is much smaller than the critical current $I_{\rm c}^{\rm eff}$ but is large enough to reach the oscillator bistability regime. Then the driven SQUID can be modeled by a Duffing oscillator. Dissipation is described as resulting from contact with a bath of harmonic oscillators. The Hamiltonian of the oscillator + bath reads
\begin{eqnarray}
\hat{H}_{\rm d}(t)&=&\hat H_S(t)+\hat H_I+\hat H_B,\label{lab}\\
\hat H_S(t)&=&\frac{\hat{p}^2}{2m}+\frac{m\omega_0^2}{2}\hat{x}^2-\gamma_S \hat{x}^4 + F(t)\hat{x},\nonumber\\
\hat H_B&=&\sum_j\hbar\omega_j\hat b_j^{\dagger}\hat b_j,\:\:\hat H_I=-\sum_j\hat{x}\lambda_j(\hat{b}_j+\hat{b}_j^{\dagger}).\nonumber
\end{eqnarray}
where $F(t)=F_0( e^{i\omega_{F} t}+ e^{-i\omega_{F} t})$ is the driving field with frequency $\omega_{F}$. The spectral density of the bath weighted with the interaction $J(\omega)=\pi\sum_j\lambda_j^2\delta(\omega-\omega_j)$ is assumed to be a smooth function for $\omega$ close to $\omega_0$. The oscillator damping constant is $\kappa= J(\omega_0)/2\hbar\omega_0m$ (cf. \cite{DK_review84}). For the case of an Ohmic bath we have $J(\omega)=2\hbar m\kappa\omega \Theta(\omega_c-\omega)$ where $\omega_c$ is a high-frequency cutoff. The interaction-induced renormalization of $\omega_0$ is supposed to be incorporated into $\omega_0$.

The steady states of the oscillator are forced vibrations. For weak damping and almost resonant driving, $|\omega_F-\omega_0|, \kappa\ll \omega_0$, and up to moderate driving amplitude $|F_0|$ these vibrations are almost sinusoidal. They can be analyzed in the rotating wave approximation (RWA) by changing to the  rotating frame with a canonical transformation $\hat U_{\rm r}(t)=\exp(-i\omega_{F} t\hat{a}^{\dagger}\hat{a})$ ($\hat{a}$ is the oscillator annihilation operator), and then by changing to dimensionless variables \cite{Serban2007,Dykman2007,Dykman1988a}
\begin{eqnarray}
\label{eq:slow_variables}
\hat x=\res(\hat Q\cos\omega_F t+\hat P\sin\omega_F t),\\
\hat p=-\res m\omega_F (\hat Q\sin\omega_F t - \hat P\cos\omega_F t),
\nonumber
\end{eqnarray}
where $\res=[2m\omega_F(\omega_0-\omega_F)/3\gamma_S]^{1/2}$.
This transformation changes phase space volumes, so that
\begin{eqnarray}
[\hat Q,\hat P]=i\lambda_S,\quad \lambda_S =\hbar\frac{3 \gamma_S}{2m^2\omega_F^2\,|\delta\omega|} , \quad \delta\omega=\omega_F-\omega_0 
\end{eqnarray}
The dimensionless parameter $\lambda_S$ plays the role of Planck's constant in these units. We assume that $\lambda_S\ll 1$, in which case the oscillator dynamics is semiclassical. The oscillator has coexisting states of forced vibrations provided $\gamma_S\delta\omega <0$.  Motivated by the description of the JBA we have chosen the {\em soft} case of the Duffing oscillator, $\gamma_S=m\omega_0^2/24 >0$, i.e., $\delta\omega<0$.
The main results can be straightforwardly generalized to the hard case by choosing $\delta\omega>0$.

If we disregard fast oscillating terms, following the RWA, the oscillator Hamiltonian $\hat H_{S{\rm r}}=\hat U_{\rm r}^{\dagger}\hat{H}_{S}\hat U_{\rm r}-i\hbar \hat U_{\rm r}^{\dagger}\partial_t\hat U_{\rm r}$ becomes time-independent,
\begin{eqnarray}
\label{eq:scaled_Hamiltonian}
&&\hat H_{S{\rm r}}=-(\hbar |\delta\omega|/\lambda_S)\hat g,\\
&&\hat g=\frac{1}{4}(\hat P^2+\hat Q^2-1)^2-\hat Q\beta^{1/2}, \nonumber
\end{eqnarray}
where $\beta$ is the dimensionless
squared amplitude of the driving field, $\beta =3\gamma_SF_0^2/2(m\omega_F\,|\delta\omega|)^3$.

For $\beta < 4/27$ the Hamilton function $H_{S{\rm r}}$ has three extremal points. In the presence of dissipation two of them, with largest and smallest $Q$, become attractors. In the laboratory frame they correspond to stable forced vibrations with amplitude $\res r_a$. The third extremum of $H_{S{\rm r}}$ is a saddle point.
For not too weak driving, the distance between the steady states in $(P,Q)$ space is large compared to the typical amplitude of quantum fluctuations $\propto \lambda_S^{1/2}$, as well as to the amplitude of classical fluctuations, which come into play for higher temperatures.

Of primary interest for us are small-amplitude fluctuations of the oscillator about steady states. They can be conveniently described by the oscillator density matrix $\hat\rho$ in the Wigner representation, $\rho_W(Q,P)=\int d\xi \exp[-i\xi P/\lambda_S]\langle Q+\xi/2\left|\hat\rho\right|Q-\xi/2\rangle$. When the oscillator is latched to one of the attractors, $\rho_W$ has a narrow Gaussian peak at that attractor. For weak damping, i.e., up to second order in the interaction $\hat H_I$ in Eq.~(\ref{lab}), and linearized close to an attractor the master equation for $\rho_W$ in the RWA reads \cite{Dykman2007}
\begin{eqnarray}
\label{eq:QKE_near_maximum}
\dot \rho_W \approx -\n(\Kb\rho_W) +\lambda_S\kappa\left(\bar n+\frac{1}{2}\right)\n^2\rho_W,
\end{eqnarray}
where we introduced vector $\Kb=\left(-|\delta\omega|\partial_Pg-\kappa Q, |\delta\omega|\partial_Qg-\kappa P\right)$
that describes the drift of the mean coordinate;
$\n=(\partial_Q,\partial_P)$ \footnote{The signs of the derivatives of $g$ in $\Kb$ are opposite to those in Ref.~\cite{Dykman2007}; also, we incorporated into $\Kb$ the factor $|\delta\omega|$}, and $\bar n=1/[\exp(\hbar\omega_F/k_BT)-1]$ is the
Planck occupation number of the oscillator. The attractor position $(Q_a,P_a)$ is found by
demanding the drift to cease, $\Kb(Q_a,P_a)={\bf 0}$, which leads to $Q_a=\beta^{-1/2}r_a^2(r_a^2-1)$,
$ P_a=-\beta^{-1/2}(\kappa/|\delta\omega|) r_a^2$, and
\begin{eqnarray}
\label{eq:equilibrium_position}
r_a^2\left[(r_a^2-1)^2+(\kappa/\delta\omega)^{2}\right]=\beta
\end{eqnarray}
($r_a^2=Q_a^2+P_a^2$) \cite{LL_Mechanics2004}.  Near attractor $a$ vectors $\Kb$ should be expanded,
\begin{eqnarray}
\label{eq:expanding_K}
\Kb(Q,P)\approx [(\Zb\n)\Kb]_a, \qquad \Zb=(Q-Q_a,P-P_a).
\end{eqnarray}
Here, $[\cdot]_{a}$ indicates that the derivatives are calculated at attractor $a$. This expansion defines a matrix $\hat{\cal K}$ with elements ${\cal K}_{nm}=[\partial_mK_n]_a$, where components 1 and 2 correspond to $Q$ and $P$, respectively.

Equation (\ref{eq:expanding_K}) is an expansion in the width of the Gaussian peak \cite{Dykman1988a}
$[\lambda_S (2\bar n+1)]^{1/2}$ as can be seen from combining Eqs.~(\ref{eq:QKE_near_maximum}) and (\ref{eq:expanding_K}) . In our units, if this quantity is much smaller than unity, the deviations of $\hat{H}_{Sr}$, Eq.~(\ref{eq:scaled_Hamiltonian}), from a quadratic form in $\hat{P}-P_a$ and $\hat{Q}-Q_a$ can be neglected. This is also reflected in the absence of higher derivatives and higher order terms in $\lambda_S$ in
Eq.~(\ref{eq:QKE_near_maximum}).


\noindent
\section{Qubit dissipation close to resonance with the oscillator}

We first consider dissipation of the qubit in the situation where its frequency is close to $2\omega_F\approx 2\omega_0$, i.e., $|2\omega_F-\omega_q|\ll \omega_F$; however, we assume that $|2\omega_F-\omega_q|$ largely exceeds the qubit dephasing rate, so that the driving does not pump the qubit resonantly. The major contribution to the correlation function $G(\omega_q)$ in Eq.~(\ref{eq:decay_excited_general}) comes from the terms in $\hat x(t_1)$ proportional to $[\hat Q(t_1)+i\hat P(t_1)]\exp(-i\omega_F t_1)$ for $t_1=t+\tau$ and $[\hat Q(t_1) - i\hat P(t_1)]\exp(i\omega_Ft_1)$ for $t_1=\tau$
(note that $\hat Q(t),\hat P(t)$ remain almost unchanged over a drive period).
We can now re-write the excursion of the quadratic term as
\begin{equation}
\delta \hat{x}^2(t)=2x_a(t)[\hat{x}(t)-x_a(t)]+[\hat{x}(t)-x_a(t)]^2.
\label{eq:decompose_noise}
\end{equation}
Here, $x_a(t)$ is given by Eq.~(\ref{eq:slow_variables}) with $\hat Q,\hat P$ replaced by $Q_a,P_a$, which have to be determined from Eq.~(\ref{eq:equilibrium_position}). With $\hat{x}(t)-x_a(t)$ being a Gauss distributed variable, Eqs.~(\ref{eq:decay_excited_general}) and (\ref{eq:decompose_noise}) give rise to two types of decay rates. One is due to the noise of $\hat{x}(t)-x_a(t)$ scaled by the prefactor $x_a$. This is a one-quantum process, where a transition between the qubit states is accompanied by creation or annihilation of one quantum of the oscillator vibrations induced by the driving field, with the energy difference compensated by the field and the thermal bath. The second is based solely on the noise of $[\hat{x}(t)-x_a(t)]^2$.
This is a two-quantum process for the oscillator, which makes transitions between the next nearest levels with energy deficit compensated also by the field and the thermal bath.  We are now going to compute both contributions.

\subsection{One-quantum noise}

We start with the first term in Eq.~(\ref{eq:decompose_noise}), the one-quantum noise, i.e., the linear correlator scaled by the attractor radius. From Eqs.~(\ref{eq:slow_variables})
\begin{eqnarray}
\label{eq:G_resonant}
&&G(\omega_q)\approx \frac{m^2\omega_F^2\delta\omega^2}{9\gamma_S^2}
r_a^2N_{+-}(\omega_q-2\omega_F),\\
&&N_{+-}(\omega)=\int\nolimits_0^{\infty}dt\exp(i\omega t)\langle \hat Z_+(t)\hat Z_-(0)\rangle_{\rm slow},\nonumber
\end{eqnarray}
where $\hat Z_{\pm}\equiv \hat Z_1\pm i \hat Z_2$ and we define $\hat Z_1 = \hat Q-Q_a, \hat Z_2=\hat P-P_a$. The averaging $\langle \cdot\rangle_{\rm slow}$ is done in the rotating frame and in the RWA. The time dependence of operators is calculated with Hamiltonian $\hat H_{S{\rm r}}+\hat H_B+ \hat U_{\rm r}^{\dagger}H_I\hat U_{\rm r}$. This Hamiltonian has the same structure (except for the explicit form of $\hat H_{S{\rm r}}$) as for a stationary oscillator with no driving, if one goes to the interaction representation via the transformation $\hat U_{\rm r}$. Then we can apply the quantum regression theorem and write \cite{DK_review84},
\begin{equation}
\label{eq:slow_correlator_defined}
\langle \hat Z_n(t)\hat Z_m(0)\rangle_{\rm slow}=\Tr \bigl[\hat Z_n\hat\rho(t|\hat Z_m)\bigr]
\end{equation}
($n,m=1,2$). The operator $\hat\rho(t|\hat Z_m)$ is the weighted density matrix; it satisfies the standard
master equation [cf. Eq.~(\ref{eq:QKE_near_maximum})], but its initial condition is $\hat\rho(0|\hat Z_m)=\hat Z_m\hat\rho^{\rm (st)}$, where $\hat\rho^{\rm (st)}$ is the RWA density matrix in the steady state.

Equation (\ref{eq:G_resonant}) reduces the problem of the qubit decay rate to calculating the oscillator power spectrum in the rotating frame. For a quantum oscillator in a given attractor it can be done using Eqs.~(\ref{eq:QKE_near_maximum}), (\ref{eq:G_resonant}), (\ref{eq:slow_correlator_defined}) and the relations for the transition to the Wigner representation
$
\left[\hat Q\hat\rho\right]_W=\left(Q+\frac{1}{2}i\lambda_S \partial_P\right)\rho_W(Q,P),\quad \left[\hat P\hat\rho\right]_W=\left(P-\frac{1}{2}i\lambda_S \partial_Q\right)\rho_W(Q,P).
$
By multiplying Eq.~(\ref{eq:QKE_near_maximum}) by $\hat Z_n\exp(i\omega t)$ and then integrating over $Q,P, t$ we obtain a linear equation for matrix $\hat N(\omega)$ of the Fourier-transformed correlators $N_{nm}(\omega)=\int\nolimits_0^{\infty} dt\exp(i\omega t)\langle \hat Z_n(t)\hat Z_m(0)\rangle_{\rm slow} $,
\begin{eqnarray}
\label{eq:N_nm_equation}
&&(i\omega\hat I+ \hat {\cal K})\hat N(\omega)=\hat {\cal N} -\frac{1}{2}i\lambda_S\hat\epsilon,\nonumber\\
&&\hat {\cal N}\hat {\cal K}^{\dagger} + \hat{\cal K}\hat {\cal N} = \lambda_S\kappa\left(2\bar n+1\right)\hat I,
\end{eqnarray}
where $\hat I$ is the unit matrix and $\hat\ep$ is the fully antisymmetric tensor (the
Levi-Civita tensor) of rank two. The matrix elements of the matrix ${\cal N}$ are ${\cal N}_{nm}=-\int dQ\,dP\,Z_nZ_m\rho^{\rm (st)}_W$. From Eqs.~(\ref{eq:G_resonant})-(\ref{eq:N_nm_equation}) one obtains for a given attractor $a$
\begin{widetext}
\begin{eqnarray}
\label{eq:resonant_power_spectrum}
{\rm Re}\,N_{+-}(\omega)=2\lambda_S\kappa
\frac{(\bar n +1)\bigl\{\left[\omega-|\delta\omega|\left(2r_a^2-1\right)\right]^2 + \kappa^2\bigr\}+|\delta\omega|^2\bar n r_a^4}
{(\omega^2-\nu_a^2)^2+4\kappa^2\omega^2},
\end{eqnarray}
\end{widetext}
where $\nu_a^2=\kappa^2+|\delta\omega|^2(3r_a^4-4r_a^2+1)$ with $r_a$ obtained from solving Eq. \ref{eq:equilibrium_position}. In the high-temperature limit $\bar n\gg 1$ Eq.~(\ref{eq:resonant_power_spectrum}) coincides with the result obtained earlier \cite{Dykman1994b} in the classical limit.

\subsection{Two-quantum noise}

The second contribution to the qubit decay rate originates from direct two oscillator quanta transitions. It is not proportional to the squared attractor amplitude $\propto r_a^2$. The decay probability is determined by the power spectrum of the squared oscillator displacement $[\hat x(t)-x_a(t)]^2$. At resonance, $|\omega_q-2\omega_0| \ll \omega_0$, the main contribution to this power spectrum comes from the Fourier-transformed correlator $\langle \hat{Z}_+^2(t)\hat{Z}_-^2(0)\rangle$. The latter was studied earlier for a nonlinear oscillator in the absence of driving \cite{DK_review84} and for linear driven oscillators \cite{Ioana07a,SerbanPRB07}. Up to moderate temperatures, $\hbar\bar n\gamma_S/m^2\omega_0^2\kappa \ll 1$, it leads to
\begin{equation}
{\rm Re}\,G(\omega_q)\approx \frac{\hbar^2}{(m\omega_0)^2}\frac{\kappa(\bar n+1)^2}{(\omega_q-2\omega_0)^2+4\kappa^2}.
\label{eq:quadrate}
\end{equation}
This term becomes important for weak driving, where the amplitude of forced vibrations is on the order of the fluctuational oscillator displacement, $r_a^2 \lesssim \lambda_S (2\bar n + 1)$. However, if the oscillator is strongly underdamped in the rotating frame, $\kappa \ll \nu_a$ (see below), the decay rate may display attractor-dependent two-quanta quasienergy resonances for $|\omega_q-2\omega_F|\sim 2\nu_a$, which will be larger than the background given by Eqs.~(\ref{eq:G_resonant}), (\ref{eq:resonant_power_spectrum}) for $\kappa r_a\lesssim \nu_a [\lambda_S (2\bar n + 1)]^{1/2}$. Both their height and position depend on the field amplitude $F_0$.

\noindent
\subsection{Discussion of the results for resonant decay}

In total, the qubit decay rate is a sum of the one-quantum, $\Gamma_{\rm 1}$, and
two-quantum rate, $\Gamma_{\rm 2}$, i.e., $\Gamma_{\rm tot}=\Gamma_{\rm 1}+\Gamma_{\rm 2}$.
These contributions are described by Eq. (\ref{eq:decay_excited_general}) with the function $G$ given by Eqs.~(\ref{eq:G_resonant}) and (\ref{eq:resonant_power_spectrum}) for $\Gamma_{\rm 1}$ and (\ref{eq:quadrate}) for $\Gamma_{\rm 2}$.

The one-quantum rate $\Gamma_{\rm 1}$ is quadratic in the forced vibration amplitude $\propto r_a$. The physical origin of this term can be understood as follows:
small fluctuations of $x(t)-x_a(t)$ are frequency-mixed with the
forced vibrations $x_a(t)$ in the laboratory frame, as described by the interaction
$\propto\delta x^2$. As a result fluctuation amplitude is multiplied by the forced-vibration amplitude, and the qubit decay rate is quadratic in the latter amplitude.

\begin{figure}[h]
\includegraphics[width=1.6in]{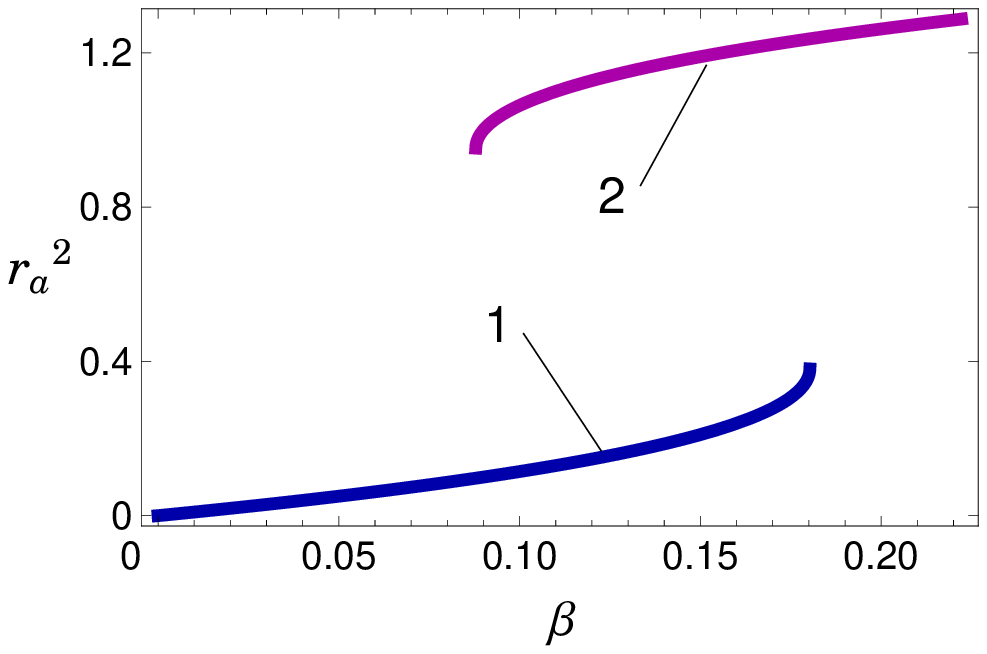}
\includegraphics[width=1.7in]{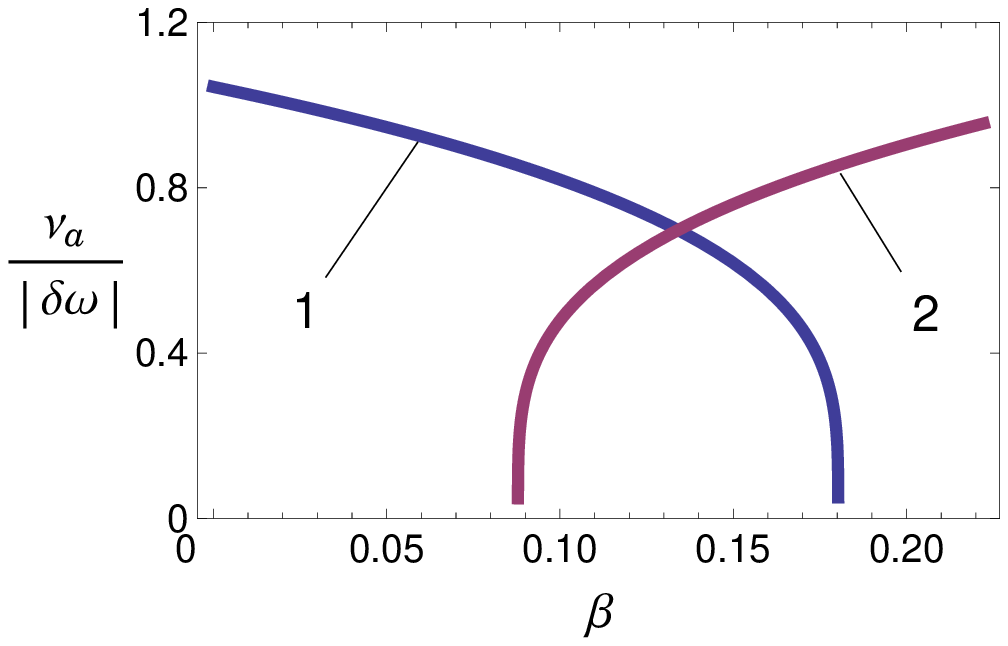}
\caption{Left panel: Squared scaled attractor radii $r_a^2$ as functions of the dimensionless field intensity $\beta$ for the dimensionless friction $\kappa/|\delta\omega|=0.3$. Right panel: The effective frequencies $\nu_a/|\delta\omega|$ for the same $\kappa/|\delta\omega$. Curves 1 and 2 refer to small- and large amplitude attractors. }
\label{fig:nu_a}
\end{figure}

Apart from the proportionality to $r_a^2$, the attractor dependence of $\Gamma_1$ is also due to the different curvature of the effective potentials around the attractors.
For weak oscillator damping $\kappa \ll \nu_a$, the parameter $\nu_a$ in Eq.~(\ref{eq:resonant_power_spectrum}) is the frequency of small-amplitude vibrations about attractor $a$. It sets the spacing between the quasienergy levels, the eigenvalues of the rotating frame Hamiltonian $H_{S{\rm r}}$ close to the attractor. The function Re~$N_{+-}(\omega)$ has sharp Lorentzian peaks at $\omega=\pm\nu_a$ with halfwidth $\kappa$ determined by the oscillator decay rate. The dependence of $\nu_a$ on the control parameter $\beta$ is illustrated in Fig.~\ref{fig:nu_a}.

The decay rate of the excited state of the qubit $\Gamma_e\propto {\rm Re}~N_{+-}(\omega_q-2\omega_F)$ sharply increases if the qubit frequency $\omega_q$ coincides with $2\omega_F\pm \nu_a$, i.e., $\omega_q-2\omega_F$ resonates with the inter-quasienergy level transition frequency. This new frequency scale results from the interplay of the system nonlinearity and the driving and is attractor-specific, as seen in Fig.~\ref{fig:nu_a}. In the experiment, for $\omega_q$ close to $2\omega_0$, the resonance can be achieved by tuning the driving frequency $\omega_F$ and/or driving amplitude $F_0$. This {\em quasienergy resonance} destroys the QND character of the measurement by inducing fast relaxation.

The analysis of the excitation rate out of the qubit ground state in the resonant case $|\omega_q-2\omega_F|\ll \omega_F$ is similar; $\Gamma_g$ is given by Eqs.~(\ref{eq:decay_excited_general}) and (\ref{eq:G_resonant}) with $N_{+-}(\omega_q-2\omega_F)$ replaced with $N_{-+}(-\omega_q+2\omega_F)$; the last function is given by Eq.~(\ref{eq:resonant_power_spectrum}) with $\omega=\omega_q-2\omega_F$ and with interchanged coefficients $\bar n+1 \leftrightarrow \bar n$.

\begin{figure}[h]
\includegraphics[width=1.6in]{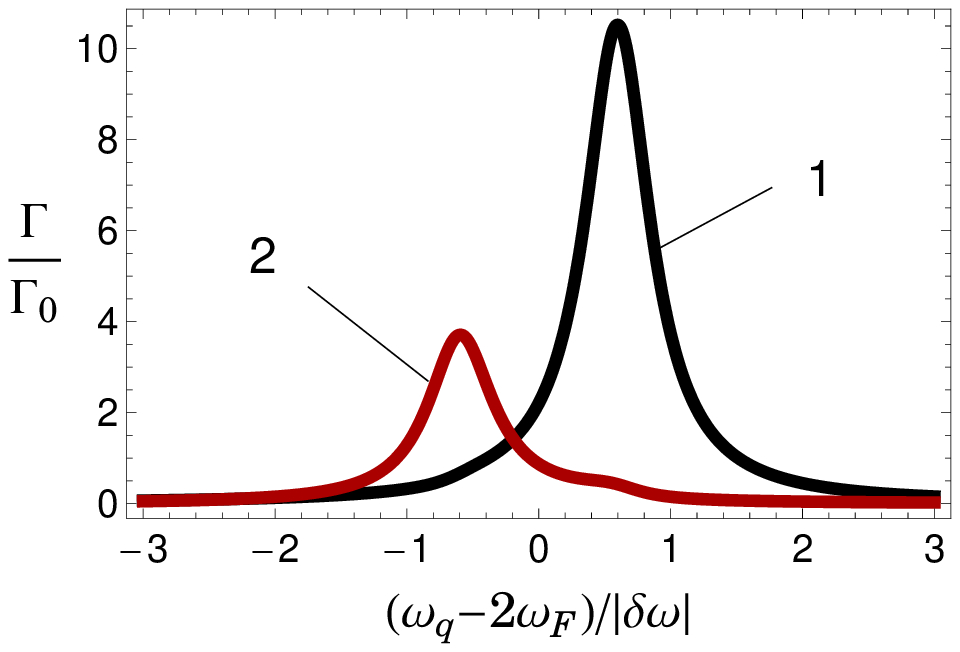}
\includegraphics[width=1.6in]{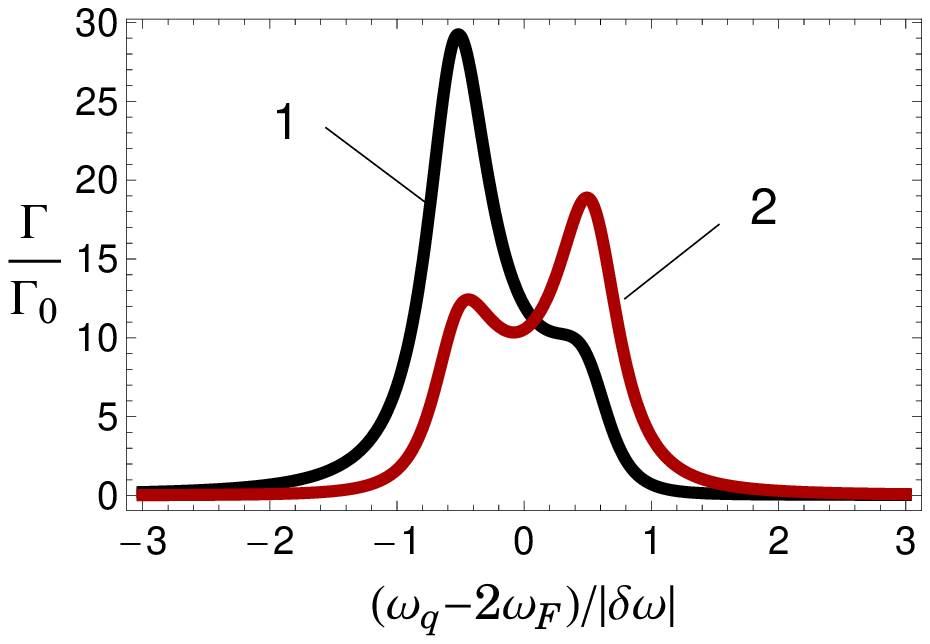}
\caption{The scaled decay rate factors for the excited and ground states, curves 1 and 2, respectively, as functions of scaled difference between the qubit frequency and twice the modulation frequency; $\Gamma_0=\hbar C_{\Gamma}r_a^2/6\gamma_S$. Left and right panels refer to the small- and large-amplitude attractors, with the values of $\beta$ being 0.14 and 0.12, respectively. Other parameters are $\kappa/|\delta\omega=0.3,\bar n=0.5$. }
\label{fig:decay_spectra}
\end{figure}

The scaled decay rates $\Gamma_{e,g}$ as functions of detuning $\omega_q-2\omega_F$ are illustrated in Fig.~\ref{fig:decay_spectra}. Even for comparatively strong damping, the spectra display well-resolved quasi-energy resonances, particularly in the case of the large-amplitude attractor. As the oscillator approaches bifurcation points where the corresponding attractor disappears, the frequencies $\nu_a$ become small (cf. Fig.~\ref{fig:nu_a}) and the peaks in the frequency dependence of $\Gamma_{e,g}$ move to $\omega_q=2\omega_F$ and become very narrow, with width that scales as the square root of the distance to the bifurcation point. We note that the theory does not apply for very small $\omega_q-2\omega_F|$ where the qubit is resonantly pumped; the corresponding condition is $(m\omega_0\Delta_q\delta C_{\rm res}^2r_a^2/\hbar\omega_q)^2\ll [T_1T_2^{-1}+ (\omega_q-2\omega_F)^2T_1T_2]$. For weak coupling to the qubit, $\Gamma_e\ll \kappa$, it can be satisfied even at resonance.

\begin{figure}[h]
\includegraphics[width=1.6in]{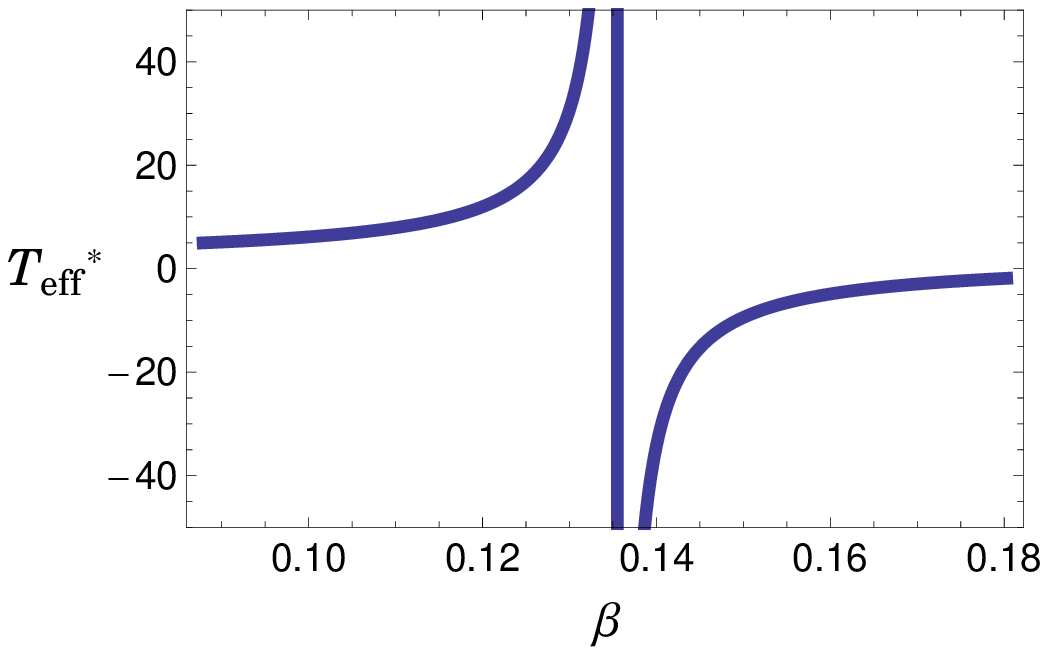}
\includegraphics[width=1.6in]{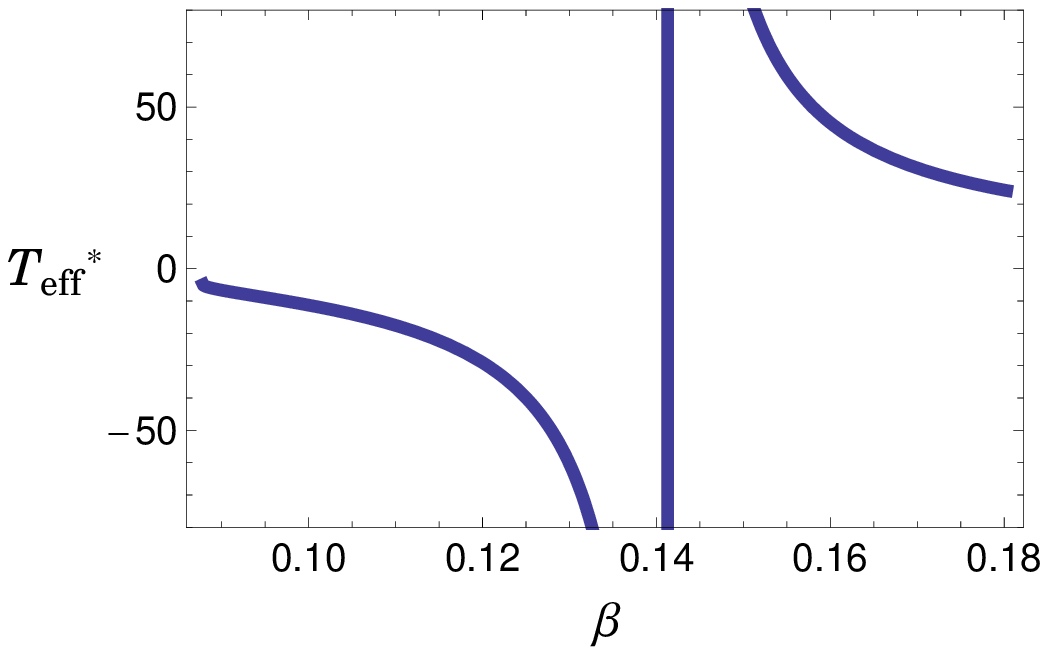}
\caption{The effective scaled qubit temperature $T_{\rm eff}^*=k_BT_{\rm eff}/\hbar\omega_q$ as function of the scaled field strength $\beta$ in the region of bistability for the small- and large-amplitude attractors, left and right panels, respectively; $(\omega_q-2\omega_F)/|\delta\omega| = -0.2$ and 0.1 in the left and right panels; other parameters are the same as in Fig.~\ref{fig:decay_spectra}.}
\label{fig:effect_temp}
\end{figure}

An important feature of the qubit relaxation in the presence of driving is that the stationary distribution over the qubit states differs from the thermal Boltzmann distribution. If the oscillator-mediated decay is the dominating qubit decay mechanism, the qubit distribution is determined by the ratio of the transition rates $\Gamma_e$ and $\Gamma_g$. One can characterize it by effective temperature  $T_{\rm eff}=\hbar\omega_q/[k_B\ln(\Gamma_e/\Gamma_g)]$. If the term in curly brackets in the numerator of Eq.~(\ref{eq:resonant_power_spectrum}) is dominating, $T_{\rm eff}\approx 2T$, but if the field parameters are varied so that this term becomes comparatively smaller $T_{\rm eff}$ increases, diverges, and then becomes negative, approaching $-2T$. Negative effective temperature corresponds to population inversion. The evolution of the effective temperature with the intensity of the modulating field is illustrated in Fig.~\ref{fig:effect_temp}.

\subsection{Linear qubit-to-oscillator coupling}

The above results can be extended also to a qubit linearly coupled to the oscillator. Such coupling was discussed in numerous contexts in the problem of a two-level system coupled to an oscillator; in particular, it underlies the broadly used Jaynes-Cummings model of the cavity quantum electrodynamics. The linear coupling leads to resonant decay of the qubit if the qubit frequency $\omega_q$ is close to the oscillator eigenfrequency, $|\omega_q-\omega_0|\ll \omega_0$ and the oscillator decay rate largely exceeds the qubit decay rate. The effect is significantly modified if the oscillator is resonantly driven. If we consider linear coupling of the Jaynes-Cummings form $V_x\sigma_xx$, the corresponding contribution to the decay rate of the excited state of the qubit is
\begin{equation}
\label{eq:linear_coupling_decay}
\Gamma_e=(V_xw/\hbar\omega_q)^2(m\omega_F|\delta\omega|/3\gamma_S)
{\rm Re} N_{+-}(\omega_q-\omega_F)
\end{equation}
(for the coupling of the form $V_z\sigma_zx$ one should replace in the above expression $V_xw$ with $V_z\delta$).

For linear coupling, the decay rate does not have the factor $r_a^2$. However, it is still different for different attractors. If the oscillator is strongly underdamped, it displays resonances whenever $\hbar|\omega_q-\omega_F|$ coincides with the oscillator quasienergy level spacing near the attractor $\hbar\nu_a$.

\noindent
\section{Qubit decay far from resonance}

For the case where $|\omega_q -2\omega_0|$ is not small compared to $\omega_0$ and $\omega_q\gg \nu_a$, the correlation function $G(\omega_q)$ can be calculated by perturbation theory to the lowest order in the oscillator-to-bath coupling. To this end we again write $\delta\hat x^2(t)\approx 2x_a(t)[\hat x(t)-x_a(t)]$. We need to find $\hat x(t)-x_a(t)$ in the time range $\sim |\omega_q-\omega_F|^{-1}$. The analysis is familiar from the studies of phonon sidebands in solids \cite{Stoneham2001}. We linearize the Heisenberg equation of motion for $\hat x(t)-x_a(t)$. The terms $\propto \gamma_S$ and $\propto F(t)$ in this equation can be disregarded, since the oscillator nonlinearity is weak, $|\gamma_S|x^2\ll m\omega_0^2$, as is also the driving; the back-action (friction) force from the bath can be disregarded as well, since $\kappa\ll |\omega_q -2\omega_0|$. However, the quantum force from the bath $\sum\nolimits_j\lambda_j(\hat b_j+\hat b_j^{\dagger})$ may have components at frequency $\omega_q\pm\omega_F$ and should be kept. Integrating the linear equation for $\hat x(t_1)-x_a(t_1)$ first with $t_1=t+\tau$ and then with $t_1=\tau$ and substituting the result into Eq.~(\ref{eq:decay_excited_general}), we obtain
\begin{eqnarray}
\label{eq:nonresonant_decay}
&&{\rm Re}\,G(\omega_q)=\frac{2\omega_F|\delta\omega|}{3m\gamma_S}
r_a^2\sum_i\frac{J(\omega_i)\Phi_T(\omega_i)}{(\omega_0^2-\omega_i^2)^2}
\end{eqnarray}
where $\omega_i$ takes on the values $\omega_q\pm \omega_F, \omega_F-\omega_q$. The function $J(\omega)$ is the density of states of the bath  weighted with the coupling to the oscillator, it is defined below Eq.~(\ref{lab}) and we assume $J(\omega)\equiv 0$ for $\omega < 0$. Function $\Phi_T(\omega_i)=\bar n(\omega_i)+1$ for $\omega_i=\omega_q\pm\omega_F$ and $\Phi_T(\omega_i)=\bar n(\omega_i)$ for $\omega_i=\omega_F-\omega_q$; here $\bar n(\omega)=1/[\exp(\hbar\omega/k_BT)-1]$. It is important that the two asymptotic expressions obtained in a different way, Eqs.~(\ref{eq:G_resonant}) and (\ref{eq:resonant_power_spectrum}) on the one hand and Eq.~(\ref{eq:nonresonant_decay}) on the other hand, coincide in the range $\nu_a,\kappa\ll |\omega_q-2\omega_F|\ll\omega_F,\omega_c$,
thus indicating that we have found the qubit relaxation rates at arbitrary frequency.

Equation (\ref{eq:nonresonant_decay}) describes decay of the qubit excitation into excitations of the bath coupled to the oscillator. This decay is mediated by the oscillator and stimulated by the driving. The decay rate is determined by the density of states of the bath at the combination frequencies $|\omega_q\pm\omega_F|$. Most notably, it is quadratic in the amplitude of the oscillator forced vibrations and therefore strongly depends on the occupied attractor.

The far-from-resonance regime is important for the experiment \cite{Picot08} with a SQUID-based bifurcation amplifier, since the amplifier was operated at a frequency far below the qubit frequency. In this experiment, the relaxation rate of the qubit for the large vibration amplitude of the oscillator was much larger than for the small amplitude, see Fig. 2c of Ref.~\onlinecite{Picot08}, and was increasing with the driving strength on the low-amplitude branch (branch 1 in the left panel of Fig.~\ref{fig:nu_a}), in qualitative agreement with the theory. It is not possible to make a direct quantitative comparison because of an uncertainty in the qubit relaxation rates noted in Ref.~\onlinecite{Picot08}; also, even though the decay rate of the oscillator  at frequency $\omega=\omega_F$ is given, the oscillator decay rate at the relevant much higher frequency $\omega=\omega_q$ is not known.

The excitation rate out of the ground state of the qubit $\Gamma_g$ is also determined by Eq.~(\ref{eq:nonresonant_decay}) with interchanged $\bar n(\omega_i) + 1
\leftrightarrow \bar n(\omega_i)$ in function $\Phi_T(\omega_i)$. If the oscillator-mediated decay is the dominating qubit decay mechanism, the effective qubit temperature $T_{\rm eff}$ depends on the interrelation between the values of $J(|\omega_q\pm\omega_F|)$. This goes beyond the temperature analysis in the
resonant case because in the former, only $J(\omega_F)$ entered into the relaxation rate whereas, for strong detuning, the environmental spectrum is probed at multiple, distinct frequencies.
The analysis here is fully similar to that of cooling and heating of oscillators \cite{Dykman1978,DK_review84}, the area that attracted much interest recently \cite{Kippenberg2008,Grajcar2008}. We note that, in contrast to the previous work, here we consider heating and cooling of a two-level system, a qubit, that results from the coupling to an oscillator rather than heating and cooling of the oscillator that may result from such coupling provided the relaxation rate of the qubit is higher than that of the oscillator.

If $J(\omega_q-\omega_F)\gg J(\omega_q+\omega_F)$, the qubit temperature $T_{\rm eff}\approx T\omega_q/(\omega_q-\omega_F)$ exceeds $T$, whereas for $J(\omega_q-\omega_F)\ll J(\omega_q+\omega_F)$ we have driving-induced cooling of the qubit, $T_{\rm eff}\approx T\omega_q/(\omega_q+\omega_F)$; for $\omega_F>\omega_q$ the temperature $T_{\rm eff}\approx - T\omega_q/(\omega_F - \omega_q)$ is negative. Heating and cooling of the qubit can be achieved also for nonresonant excitation of the oscillator \cite{Dykman1978,DK_review84,Kippenberg2008}.

For weak driving and $\omega_q$ far from $2\omega_0$, qubit decay may be due to two-quanta transitions in which an oscillator makes a transition between its neighboring levels and a quantum of the bath is created or annihilated (for stronger oscillator-bath coupling, two bath quanta transitions may be also important). The corresponding rate is determined by ${\rm Re}\,G(\omega_q)\approx (2\hbar/m^3\omega_0)\sum\nolimits_i J(\omega_i) (\omega_0^2-\omega_i^2)^{-2} \Phi_T(\omega_i)\Phi_{Ti}$, where $\omega_i=\pm\omega_q \pm\omega_0$. Here, function $\Phi_T(\omega_i)$ is the same as in Eq.~(\ref{eq:nonresonant_decay}) (except that $\omega_F$ should be replaced with $\omega_0$), whereas $\Phi_{Ti}=\bar n+1$ for $\omega_i=\pm (\omega_q-\omega_0$ and $\Phi_{Ti}=\bar n$ for $\omega_i=\omega_q+\omega_0$. This expression coincides with Eq.~(\ref{eq:quadrate}) for $\omega_0 \gg |\omega_q-2\omega_0| \gg \kappa$.

The results on nonresonant oscillator-mediated decay can be extended also to the case of linear qubit-oscillator coupling, with coupling energy $V_x\sigma_x x$. In this case $\Gamma_e=2(V_xw/\hbar m\omega_q)^2(\omega_q^2-\omega_0^2)^{-2}J(\omega_q)[\bar n (\omega_q)+1]$. The decay rate is independent of the attractor occupied by the oscillator.

Due to the factor $(w/\delta)^2$ in $T_2^{-1}$, i.e. the particular choice of working point, the dephasing rate of the qubit is several orders of magnitude larger that the relaxation rate, allowing for a good, near-QND, qubit detection.

\section{Conclusion}

In conclusion, the presented approach describes the relaxation of a qubit coupled to an underdamped driven quantum oscillator, both in the regime where the qubit transition frequency $\omega_q$ resonates with the oscillator frequency or its second overtone and where it is far away from resonance. The proposed mechanism shows to what extent quantum measurements with bifurcation amplifiers are of the nondemolition type.

The qubit relaxation rate is expressed in terms of the power spectrum of the oscillator. We find this power spectrum in the explicit form near resonance where, as we show, it can have a double-peak structure as function of frequency. This result and the proposed method are general for a nonlinear oscillator, and thus go beyond the problem of qubit relaxation.

Both in the resonant and nonresonant regimes the qubit decay rate strongly depends on the attractor the oscillator is latched to. Far from  resonance, including the experimentally important range of a high qubit frequency, and for the coupling quadratic in the oscillator coordinate the qubit decay rate is quadratic in the amplitude of the oscillator's forced vibrations. For $\omega_q$ close to $2\omega_0$, the decay rate is resonantly enhanced. For a strongly underdamped oscillator it displays narrow maxima as a function of the control parameters once $\hbar\omega_q$ goes through the appropriate oscillator quasienergy, which depends on the attractor. We observe that the decay mediated by a driven oscillator changes the qubit temperature, leading to heating, cooling, or population inversion depending on the frequency detuning. In the case of linear qubit-oscillator coupling, the qubit relaxation rate does not contain the squared vibration amplitude as a multiplying factor, but for $\omega_q$ close to $\omega_0$ it displays attractor-dependent quasienergy resonances.

IS and FKW were supported through NSERC discovery grants and quantumworks as well
as the EU through EuroSQIP and the DARPA through the Quest program. MID was supported in part by the NSF grant  EMT/QIS-0829854.


\end{document}